\documentclass{webofc}
\usepackage{subfigure}
\usepackage[varg]{txfonts}   
\usepackage{natbib}
\begin{document}
\title{Physics Validation of Novel Convolutional 2D Architectures for Speeding Up High Energy Physics Simulations}
%
%

\author{\firstname{Florian} \lastname{Rehm}\inst{1,2}\fnsep\thanks{\email{florian.matthias.rehm@cern.ch}}, 
\firstname{Sofia} \lastname{Vallecorsa}\inst{1}, 
\firstname{Kerstin} \lastname{Borras}\inst{2,3} and 
\firstname{Dirk} \lastname{Krücker}\inst{3}
}

\institute{CERN, Esplanade des Particules 1, Geneva, Switzerland 
\and
           RWTH Aachen University, Templergraben 55, Aachen, Germany
\and
           DESY, Notkestraße 85, Hamburg, Germany
          }

\abstract{ 
The precise simulation of particle transport through detectors remains a key element for the successful interpretation of high energy physics results.
However, Monte Carlo based simulation is extremely demanding in terms of computing resources. This challenge motivates investigations of faster, alternative approaches for replacing the standard Monte Carlo approach.

We apply Generative Adversarial Networks (GANs), a deep learning technique, to replace the calorimeter detector simulations and speeding up the simulation time by orders of magnitude. We follow a previous approach which used three-dimensional convolutional neural networks and develop new two-dimensional convolutional networks to solve the same 3D image generation problem faster. Additionally, we increased the number of parameters and the neural networks representational power, obtaining a higher accuracy. We compare our best convolutional 2D neural network architecture and evaluate it versus the previous 3D architecture and Geant4 data. Our results demonstrate a high physics accuracy and further consolidate the use of GANs for fast detector simulations.
}
\maketitle

\section{Introduction}
\label{sec:introduction}
Accurate Simulations of elementary particles in High Energy Physics (HEP) detectors are fundamental to correctly reproduce and interpret the experimental results.
Detector simulations rely on Monte Carlo-based methods, such as the Geant4 toolkit \cite{Geant4} which suffer from high computational costs: currently more than half of the Worldwide Large Hadron Collider (LHC) Grid resources are used for the generation and processing of simulated data \cite{RoadmapHEP}. 
Unfortunately, the operational requirements related to the future High Luminosity phase of the LHC will exceed the expected available computational resources drastically even if taking technological improvements into account \cite{HL-LHC}.
The more excessive need for simulation capacities motivates the research on alternative fast simulation approaches. Based on generative models several prototypes have demonstrated great potential \cite{de2017learning,paganini2017calogan,Salamani2018,dijet,gan_lhcb}. While orders of magnitude speed ups are promised, the most crucial goal is given by the high level of accuracy required in order to correctly reproduce the particle showers.

Generative Adversarial Networks (GANs) represent one promising alternative  approach which have been employed primarily for simulating calorimeter detectors \cite{EnergyGAN}. 
In this paper we introduce three novel GAN architectures which use Convolutional Neural Networks (CNNs) in two-dimensional (Conv2D) layers instead of convolutional three-dimensional (Conv3D) layers for representing 3D images. The baseline model to which we compare is taken from Ref.  \cite{EnergyGAN}. Due to the large computational complexity of three-dimensional convolutions with respect to the two-dimensional case, we expect to observe a speed up and a reduction of the necessary computational resources in terms of needed memory. Additionally, we increased the number of parameters of the Conv2D neural networks what leads to a higher representation power and therefore to a higher physics accuracy. 
In the end we select the best among our new Conv2D architectures and make a detailed physics comparison to the baseline Conv3D model and to Geant4 data.

The following section 2 provides a brief overview on related work. The GAN approach for calorimeter simulations is introduced in section 3. In section 4, we explain the novel Conv2D architectures, while section 5 describes the evaluation of the results of our tests in terms of computational resources and physics accuracy. The last section summarizes the conclusion.

\section{Related Work}
\label{sec:relatedwork}
The first successful application of CNNs to image recognition problems dates back to the 1990s with the LeNet architecture designed to classify hand-written digits \cite{LeNet}. In 2014, VGGNet \cite{VGG} was introduced: with 140M parameters it proved for the first time, that the network depth remain a critical aspect to achieve good performance. 
CNNs are extremely successful and today employed in a broad range of various applications such as image classification and object tracking: CSPDarkNet-53 \cite{YOLOv4} is a recent example for object detection with YOLOv4; ResNet50 \cite{ResNet} and Inception-v4 \cite{Inception_v4} are optimised for 2D image classification. CNNs play today a key role in many research areas and applications, ranging from medical imaging to autonomous driving \cite{GANS_Applications}.

On the other hand, the use of 2D convolutions to solve three-dimensional problems is however limited. In \cite{2Dslicemethod} a classification of a 3D shape is achieved by building 3 two-dimensional slices along the three shape's canonical axes. Three independent images are created and classified by another neural network. \cite{conv2duse} expanded on the previous method by stacking a set of Conv2D layers along the $x$-, $y$- and $z$- axes, thus rebuilding a 3D image from the initial set of 2D image slices. This approach resulted in higher memory efficiency.

Within the field of High Energy Physics simulations, \cite{Torben_Thesis} proposed a network consisting of Conv2D layers to generate the three dimensional volume of a calorimeter: a three-dimensional 12x15x7 image is generated by stacking 12 two-dimensional 15x7 images together. This model yields accurate physics results and an impressive $6\,660$x speed up with respect to standard Monte Carlo techniques, running on a NVIDIA GTX 1080 card. We compare our results to the architecture presented in \cite{Torben_Thesis}.  

\section{Calorimeter Simulation with GANs}
\label{sec:calorimeters}

Generative Adversarial Networks (GANs) consist of two deep neural networks trained by employing an adversarial approach. The first network, the generator, generates images (fake images) starting from a latent vector. The second network represents the discriminator, which evaluates and distinguishes the generated images (fake images) from the true images, belonging to the training set. The GAN training is successful, when the discriminator classification probability approaches 50\% for both, the true and the fake images.
Interpreting the calorimeter output as a three-dimensional image encourages the use of architectures consisting primarily of convolutional layers. 
The model developed in \cite{EnergyGAN}, which uses Conv3D layers, is building the baseline in this study and comprises 3D images of 25x25x25 calorimeter cells (pixels).
An accurate representation of the energy pattern across the pixels is key, since this pattern carries physics information used to reconstruct the particles' properties, such as their type.

Therefore, the generator has to reliably reproduce the particle shower development along all three volume axes. 
In the following section are the details of the alternative network approaches which we have developed and analysed introduced. Additionally, we discuss and evaluate the corresponding results by comparing them to the baseline Conv3D architecture and to Geant4.

\section{New Convolutional 2D GAN Architectures}
\label{sec:Conv2D}
\begin{figure*}[ht!]  
    \centering
    \includegraphics[width=.9\textwidth, clip=true]{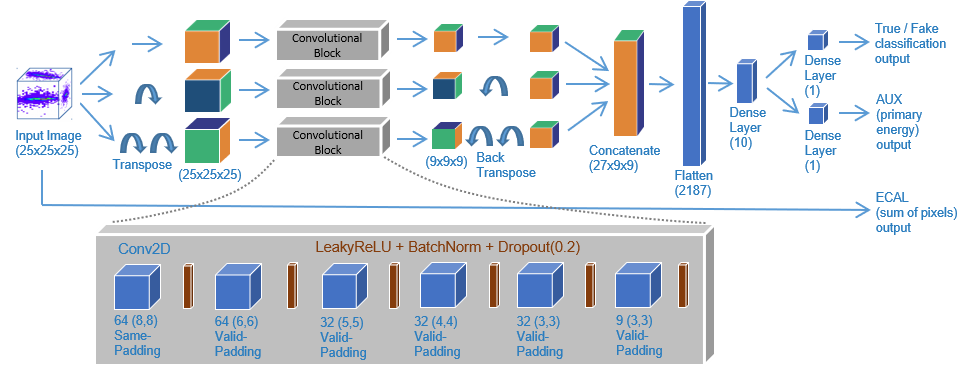}
    \caption{New Conv2D discriminator network used for training all generator networks.}
    \label{fig:Conv2Ddisc}
\end{figure*}

We developed three generator architectures and one discriminator architecture using Conv2D layers. It should be noted that each of the three Conv2D GAN models is built using the same topology for the discriminator network in order to efficiently compare the representational power of the different Conv2D generators to each other.
The Conv2D discriminator is presented in figure \ref{fig:Conv2Ddisc}.
Its inputs are either generated images (fake images) or training images (true images). The discriminator outputs three values: the first is the typical GAN true/fake probability \cite{goodfellow} which is used to calculate a binary cross entropy loss \cite{crossentropy}. The second loss (named AUX, for AUXiliary loss) represents the result of a regression task on the initial particle energy $E_p$, that the discriminator estimates from the images using a dense layer. It is implemented as a Mean Absolute Percentage Error (MAPE) \cite{MAPE}. 
The third discriminator output comes from a lambda layer, calculating the sum over the pixels of the input image which, therefore, corresponds to the total energy of the input image. It is entitled ECAL and uses again the MAPE loss function.

\subsection{The G1 Generator: The Three Branches Architecture}
\begin{figure*}[ht!]  
    \centering
    \includegraphics[width=.95\textwidth, clip=true]{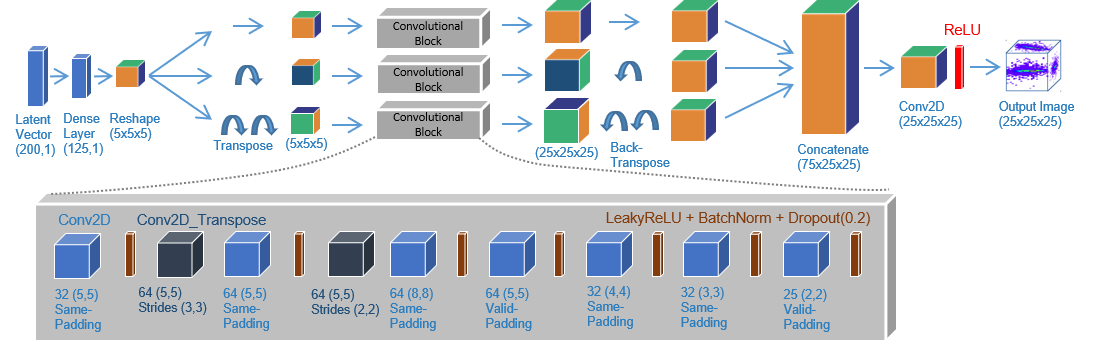}
    \caption{The G1 generator: Three branches architecture.}
    \label{fig:gen_3path}
\end{figure*}

The G1 generator implements a three-branches architecture similar to the one used for the discriminator. It outputs is a 3D 25x25x25 pixels image. The latent space is initialised as a set of 200 random numbers, drawn from a uniform distribution, and it is multiplied by the primary particle energy $E_p$. In addition to the Conv2D layers, the generator network includes transposed 2D convolutional (Conv2D\_transpose) layers to increase the image size, batch normalization (BatchNorm), rectified linear units activation function (ReLU), linear ReLU activation functions (LeakyReLU) and dropout layers (Dropout).
The initial latent vector, reshaped to a 3D grid, is transposed along the three axes and run through the three Conv2D blocks. After transposing it back in the opposite direction, the three branches outputs are concatenated and input to a final Conv2D layer in order to achieve the desired dimensions (25x25x25). These transpositions allow the Conv2D network to learn correlations within all three image dimensions. The G1 generator architecture is shown in figure \ref{fig:gen_3path}.

\subsection{The G2 Generator: The 25 Branches Architecture}
\begin{figure*}[ht!]  
    \centering
    \includegraphics[width=.95\textwidth, clip=true]{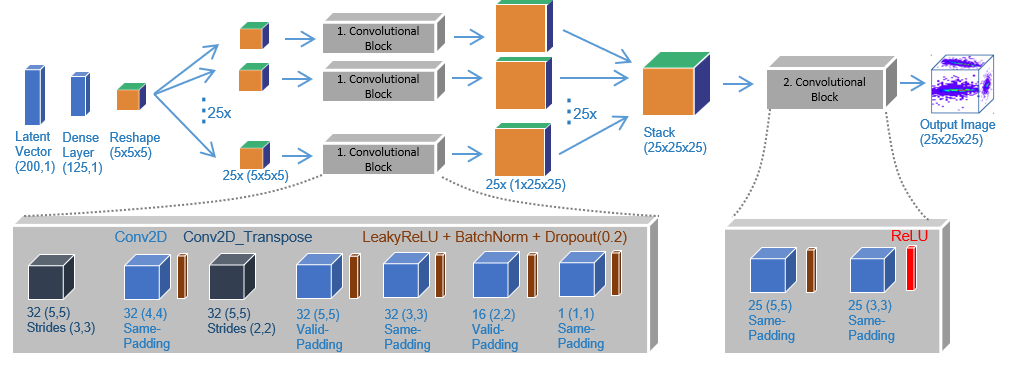}
    \caption{The G2 generator: 25 branches architecture.}
    \label{fig:gen_25path}
\end{figure*}

The second generator architecture (G2) follows the structure outlined in section \ref{sec:relatedwork}. It consists of 25 branches, one for each of the 25 layers along the $z$ axis. The 25 branches create 25 two-dimensional images (with a 25x25 shape), that are stacked in order to rebuild a 3D 25x25x25 volume. Finally, two additional Conv2D layers are applied to allow some interaction between the single dimensions. The G2 generator is indicated in figure \ref{fig:gen_25path}. 

\subsection{The G3 Generator: The Single Branch Architecture}
\begin{figure*}[ht!]  
    \centering
    \includegraphics[width=.95\textwidth, clip=true]{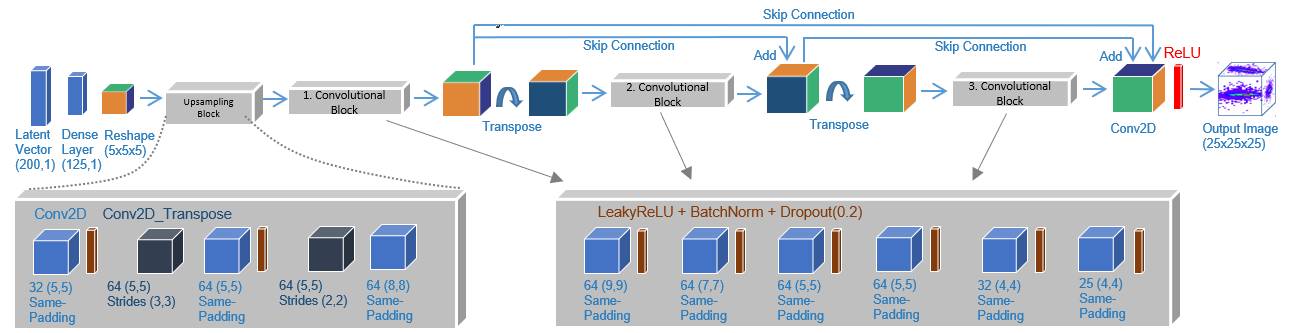}
    \caption{The G3 generator: Single branch architecture with three skip connections.}
    \label{fig:gen_1path}
\end{figure*}

The third architecture we developed and tested, the G3 generator, is a much deeper network. It alternates Conv2D layers to transpositions along the three axes, in such a way that each axis is represented once as the convolutional layers channel dimension. In order to improve convergence in the training, we have introduced skip connections across the layers, as shown in figure \ref{fig:gen_1path}. Due to its depth, the G3 model training process requires a large number of epochs before reaching convergence.

Table \ref{tab:InferenceTimes} summarizes the total number of parameters for the three architectures: the initial Conv3D network has less than half the number of parameters and a smaller number of convolutional layers, compared to the three Conv2D architectures. 
Furthermore, the Conv3D model consists of most of its parameters concentrated in the first dense layer (630k out of 753k parameters $\hat{=}$ 84\%). However, for image processing convolutional layers typically tend to have a higher representational power than dense layers while being able to keep the number of parameter smaller. Moreover, taking into account the Conv3D to Conv2D differences in terms of computational resources, we expect to be able to decrease computation time by converting neural network architectures from Conv3D to Conv2D layers. This we expect even though the model in \cite{EnergyGAN} contains only 4 Conv3D layers compared to the 28, 177, and 25 layers in the G1, G2 and G3 architectures.

\section{Evaluation}
\label{sec:evaluation}
In this section we evaluate the performance of the three Conv2D generators. We assess in terms of computational resources required by the inference and training processes and the physics accuracy of the generated images. We compare to the initial Conv3D network architecture and to Geant4.

\subsection{Computational Evaluation}
In order to measure the inference time, we process an initial set of 20 warm-up batches and then use 100 inference steps including 20 batches for each of the measurements.
Tests were performed on a Nvidia Tesla T4 GPU using Python version 3.6.8, TensorFlow version 2.2.0 and a batch size of 128. Table \ref{tab:InferenceTimes} summarises the inference results for each model. 
\begin{table}[h!]
 \centering
 \caption{Summarizes the number of parameters and the number of convolutional layers (Conv) for each generator model. Additionally, the inference times on the GPU are shown. The speed up is given with respect to Geant4 and in bracket versus the baseline Conv3D network. The last row shows the average GPU utilization during inference.}
 \begin{tabular}{c c c c c c } 
 \hline
 Model & Parameters &  Conv  & Inference  & Speed Up & Utilization \\ [0.5ex] 
 \hline
 Conv3D &	752k  &  $4$  &	7.00\,s & 	$6\,200$x (1.00x)  &	78.75\% \\ 
  \hline
 G1 3-branches &	$2\,052$k & $28$  &	4.91\,s &	$8\,000$x (1.29x)  &	21.75\%    \\  
  \hline
 G2 25-branches & $2\,131$k &  $177$  & 17.48\,s &	$2\,480$x (0.40x) & 21.83\%   \\ 
  \hline
 G3 1-branch &	$2\,397$k & $25$   &	6.21\,s &	$7\,006$x (1.13x)  &	32,90\%   \\  
 \hline
\end{tabular}
 \label{tab:InferenceTimes}
\end{table}

The values clearly demonstrate, that the G1 3 branches architecture has the lowest inference time and therefore the highest speed up (factor of 1.29x) although it has more than twice as much parameters as the Conv3D architecture. It should be noted, that the G1 architecture has an enormous $8\,000$x speed up versus the Geant4 simulation, aimed at being replaced by a more efficient GAN model.
Additionally, the Conv2D G1 architecture has a considerably lower GPU utilization compared to the Conv3D baseline model (factor of 3.6x). With running multiple streams on the GPU the inference time could be further decreased by a potential factor of 4.64x ($=1.29 \cdot 3.6$) compared to the baseline Conv3D model. 
The G2 model exhibits a higher inference time than the Conv3D model, potentially caused by the larger number of convolutional layers. However, it shows the same low GPU utilization as the G1 architecture.
The G3 model has a marginally higher inference time as the Conv3D model, but considerably less GPU utilization.

Using the same hardware, we have investigated the training time: results per epoch are presented in table \ref{tab:TrainingTimes}. The Conv3D network is by far the slowest and the G1 3-branches network the fastest, demonstrating a speed up of 6.5x. The G2 25-branches network is the slowest Conv2D model.

\begin{table}[h!]
 \centering
 \caption{Training time per epoch for the different models on a Nvidia Tesla T4 GPU and the corresponding speed up compared to the Conv3D architecture. Additionally shown, are the MSE validation index for all GAN models compared to Geant4 simulation. The lower the MSE value, the better the physics accuracy and representation.}
 \begin{tabular}{c c c c} 
 \hline
 Model & Time per Epoch [min] & Speed Up & MSE \\ [0.5ex] 
 \hline
 Conv3D &	258 &	 1x   &	0.065      \\ 
  \hline
 G1 3-branches &	40 &	6.5x   &	0.027   \\ 
  \hline
 G2 25-branches & 71 &	3.6x     & 0.048 \\ 
  \hline
 G3 1-branches &	47 &	5.5x   &	0.071    \\ 
 \hline
\end{tabular}
 \label{tab:TrainingTimes}
\end{table}

\subsection{Physics Evaluation}
\begin{figure*}[ht!]  
    \centering
    \includegraphics[width=.9\textwidth]{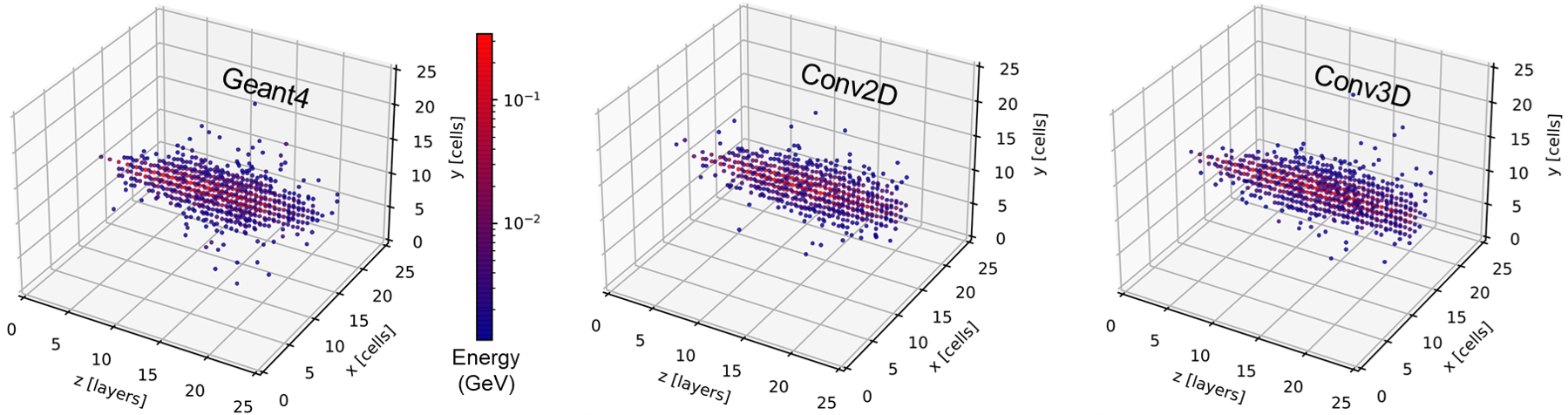}
    \caption{Example Geant4 (left), Conv2D generator (middle), Conv3D generator (right) 3D images for a 400 GeV electron. The primary particle enters the calorimeter from the left in the middle of the $x$- and $y$-axis ($x=y=13$) at $z=0$.}
    \label{fig:example_showers}
\end{figure*}

Evaluating the performance of a generative model for image generation remains a non-trivial task which is mainly done with visually inspections of the energy patterns across the calorimeter volume and comparison to Monte Carlo predictions.
An initial visual inspection demonstrates that the Conv2D and the Conv3D model can produce realistic images, similar to Geant4 (see figure \ref{fig:example_showers}).

In order to quantify the agreement of the GAN models with Geant4, we define a composite accuracy value. It is calculated by building 2-dimensional projections of the particle shower distributions along the $x$-, $y$- and $z$-axis (averaged over $20\,000$ samples) for the GAN models and the Monte Carlo. We measure the total mean squared error (MSE) between Geant4 and the corresponding GAN model (summarised in table \ref{tab:TrainingTimes}). 
The G1 architecture shows the lowest $\mathrm{MSE}=0.027$ which corresponds to the highest accuracy of our models. The G2 architecture has as well a lower $\mathrm{MSE}=0.048$ compared to the Conv3D model with a $\mathrm{MSE}=0.065$. The G3 model performs worst with a $\mathrm{MSE}=0.071$. 
We compared our Conv2D models with different metrics and in all of them the G1 architecture performs best. Therefore, we focus in the following exclusively on the G1 architecture and compare it to the baseline Conv3D model and Geant4.

\begin{figure}[ht!]  
    \centering
        \includegraphics[width=.8\textwidth, clip=true]{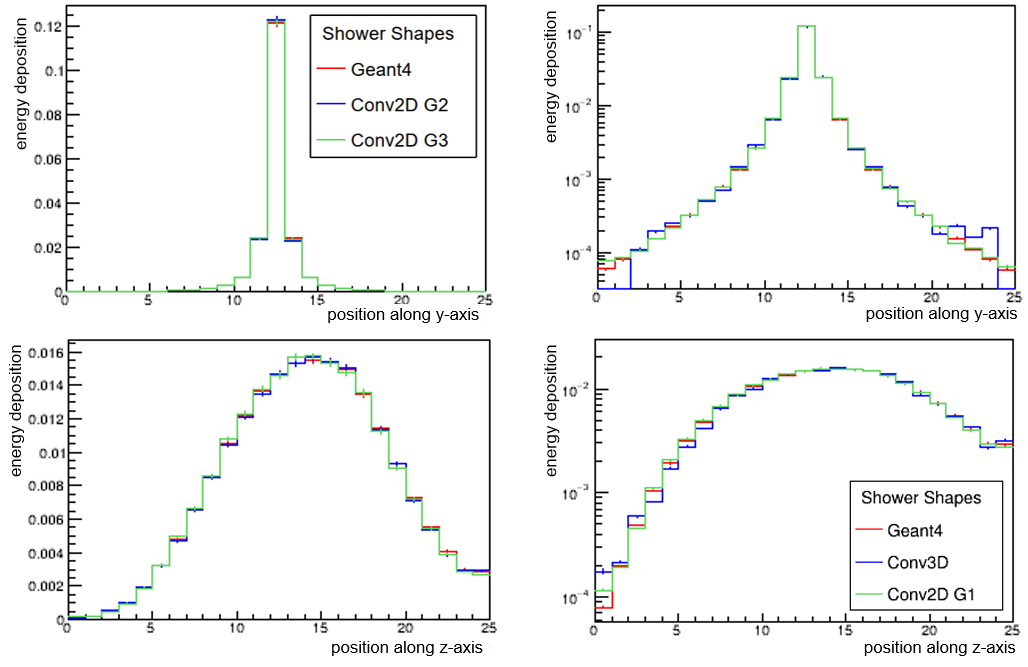}
    \caption{Particle shower distributions along the $y-$ (top) and $z-$  (bottom) axes in linear (left) and logarithmic (right) scale. The Geant4 prediction, the Conv3D model and the Conv2D G1 model outputs are shown in red, blue and green respectively.}
    \label{fig:hist0}
\end{figure}

2D projections of the energy shapes along the $y$- and $z$- axis (averaged over $20\,000$ samples) are presented in figure \ref{fig:hist0} in linear (left) and logarithmic (right) scale. The $x$-axis projections are similar to the ones of the $y$-axis and therefore not shown. In general the GAN models are close to Geant4. A more detailed analysis of the logarithmic $y$-axis distribution (top right), shows that the tails of the Conv3D distribution (in blue) are off, similar as it was already observed in \cite{EnergyGAN}, while the Conv2D model (green) improves the description of the tails.

\begin{figure}[ht!]  
    \centering
    \subfigure{\includegraphics[width=.45\textwidth, clip=true]{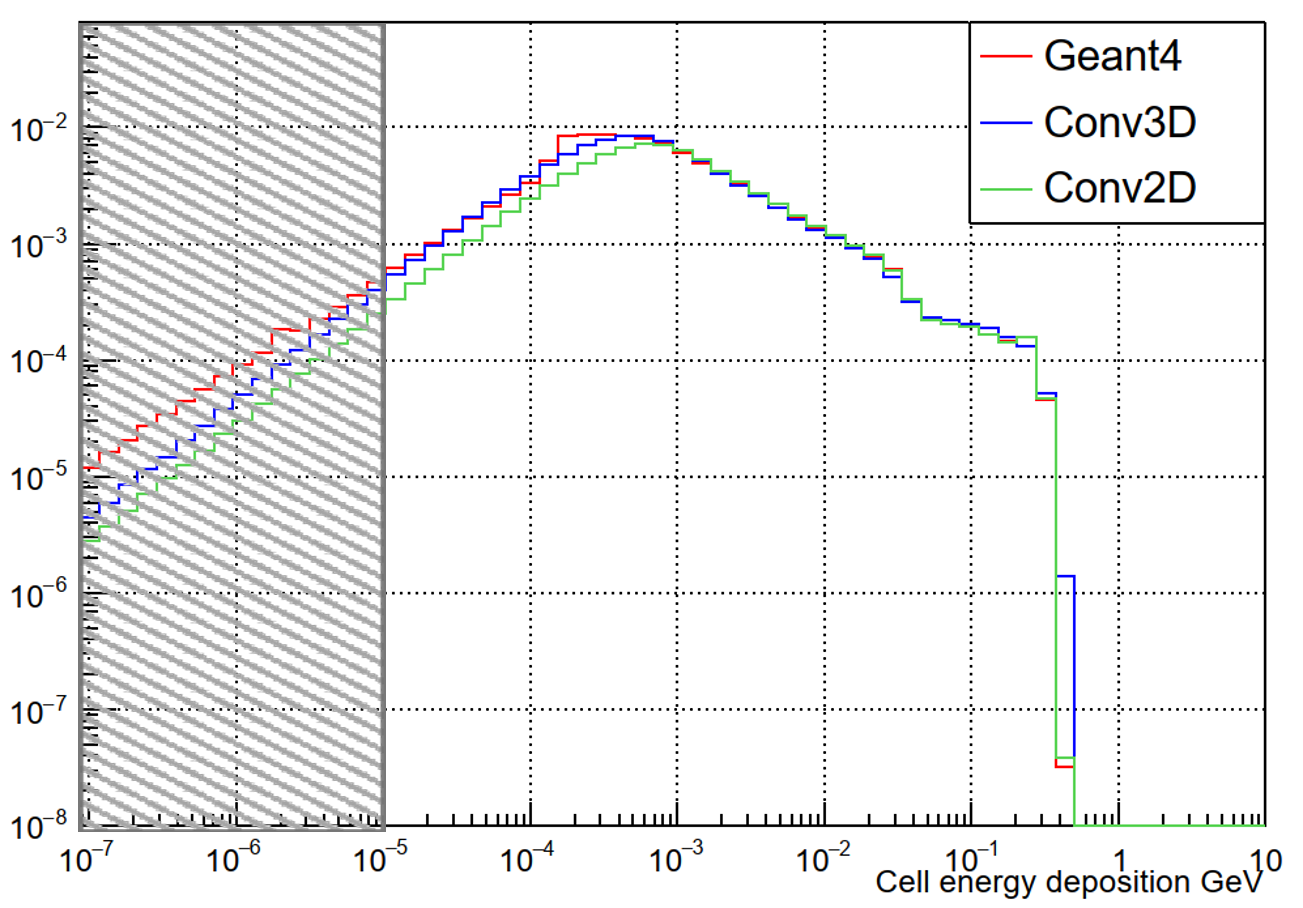}}
    \subfigure{\includegraphics[width=.45\textwidth, clip=true]{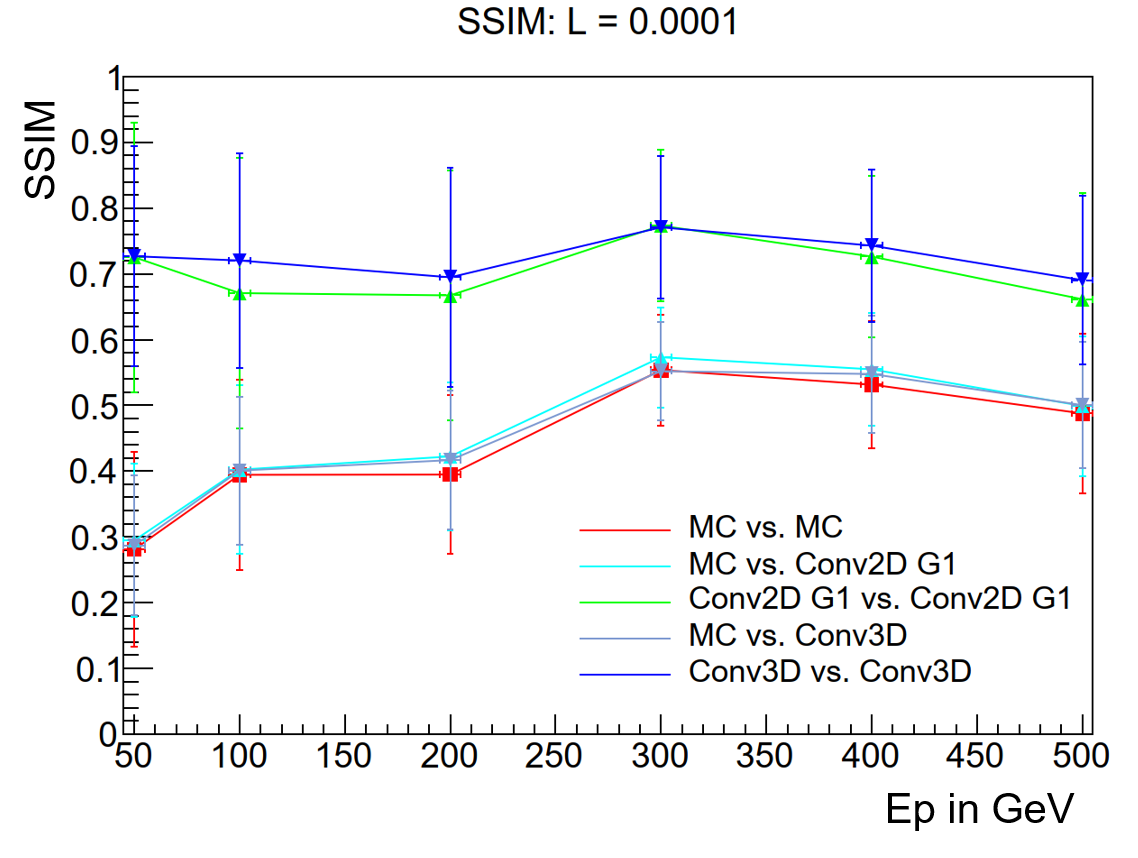}}
    \caption{Single cell energy deposition (left). Calibrated SSIM index  ($L=10^{-4}$) (right), see text for further details.}
    \label{fig:flatecal_ssim}
\end{figure}

The left panel in figure \ref{fig:flatecal_ssim} shows the single cell energy deposition for $20\,000$ showers with input energies within the full energy range of 2-500 GeV. We observe, that the high energy hits are well described by both models, even though below 1 MeV both GAN models fail to correctly describe the expected distribution. 
For this reason, we apply a minimal energy threshold of $10^{-5}$ GeV (grey shaded region) for all further validation steps.

In figure \ref{fig:flatecal_ssim} on the right we show the Structural Similarity Index (SSIM) \cite{SSIM} between different image sets. The SSIM measures the perceptual difference between similar images but does not judge if one image is better than the other.
The SSIM index is sensitive to the dynamic range of the pixels and must be adjusted using the parameter $L$ in order to maintain sensitivity. We have scanned several $L$ values ranging over orders of magnitude from $L=0.1$ to $L=10^{-6}$. We found that the SSIM index stabilizes starting from $L=10^{-4}$ and below. 
We can see in figure \ref{fig:flatecal_ssim} (right) that the SSIM values for Monte Carlo (MC) of the Geant4 tool vs. MC, MC vs. G1 and MC vs. Conv3D are close to each other, proving that the similarity between GAN generated images and Monte Carlo images is of the same level as the similarity measured within Monte Carlo images. However, the generated images vs. generated images SSIM is clearly higher proving that both GAN models generate images that are more similar to each other than what is predicted by Monte Carlo. This result highlights a well known weakness of GAN models which tend to produce samples exhibiting a lower diversity than the original data set. While we plan to continue our investigation in order to better quantify GAN performance in terms of support space size, sample diversity and mode dropping effects, the current result is well within the precision level we expect of fast simulation models.

\begin{figure}[ht!]  
    \centering
    \subfigure{\includegraphics[width=.45\textwidth, clip=true]{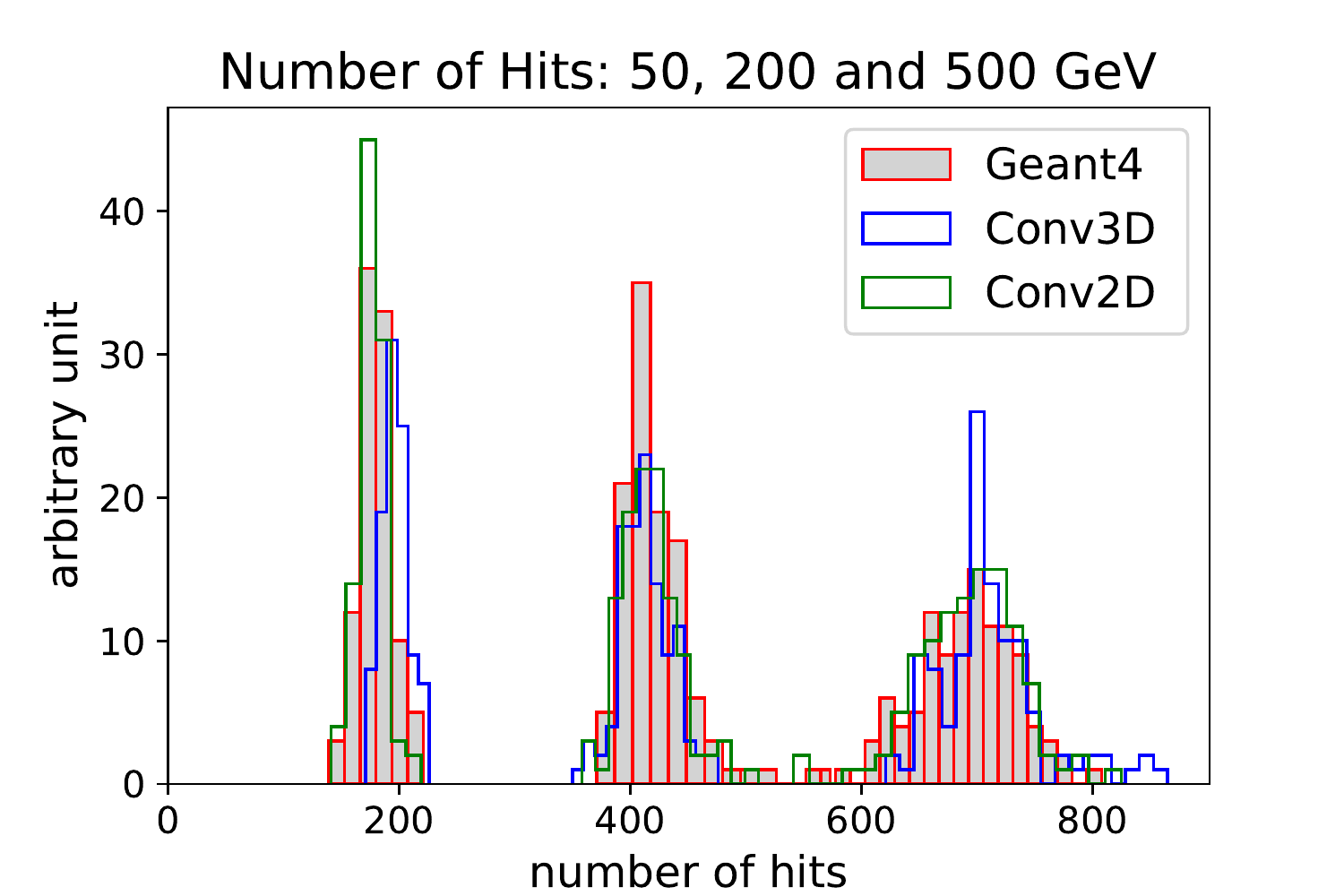}}
    \subfigure{\includegraphics[width=.45\textwidth, clip=true]{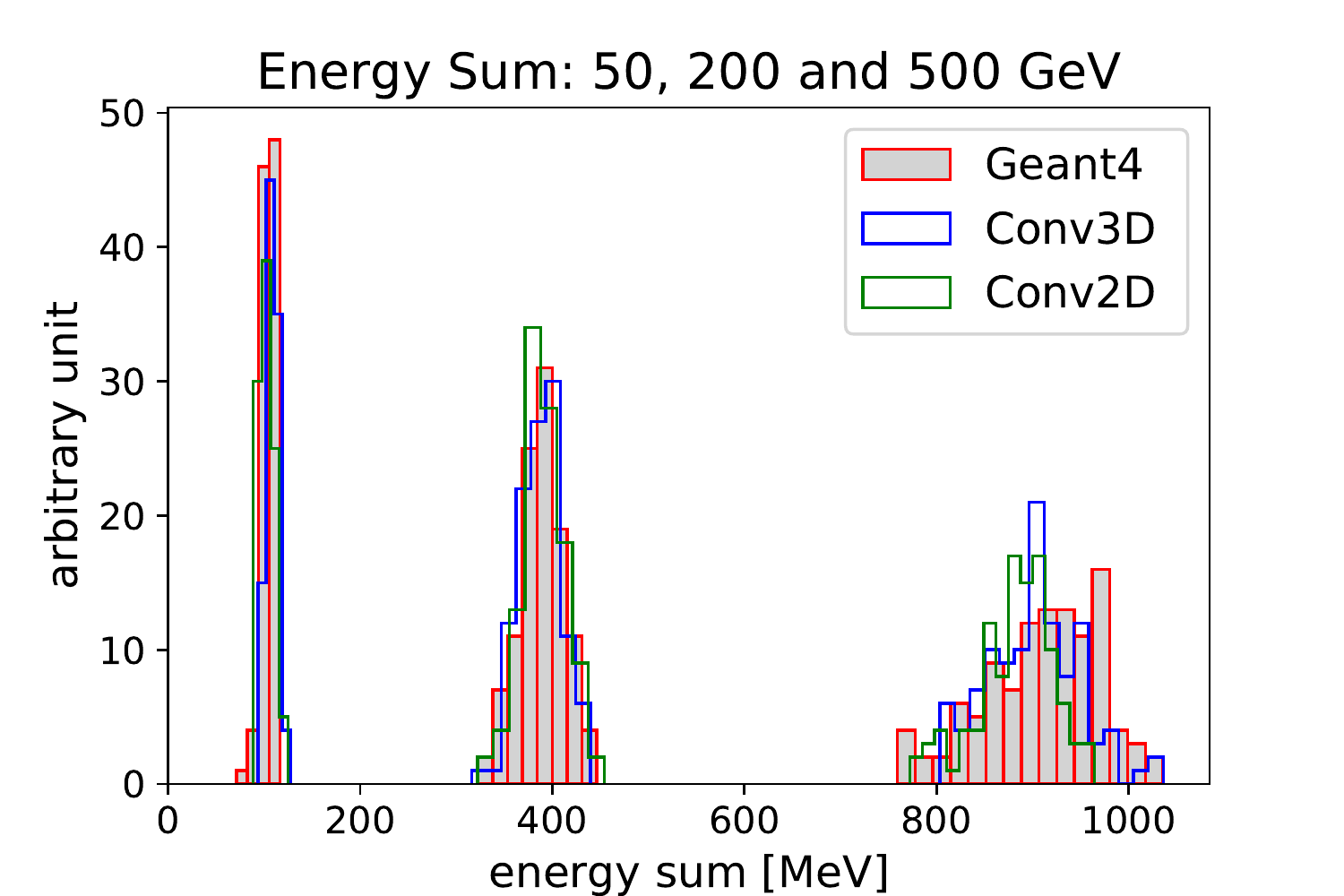}}
    \caption{ (left) Number of hits. (right) Total deposited energy. A  $10^{-5}$ GeV single cell energy threshold is applied.}
    \label{fig:number_of_hits}
\end{figure}

In an effort to standardise a performance evaluation of generative models applied to calorimeter simulations, we have chosen the following validation plots following the format proposed in \cite{Buhmann:2020pmy}.
Figure \ref{fig:number_of_hits} shows on the left side the total number of hits and on the right side the total deposited energy, for three input particle energies (50, 200 and 500 GeV). In our case, the generation energy spectrum is continuous, so in order to probe how the number of hits varies with the input particle energy we build  $\pm 2$ GeV bins around the corresponding central energy value: the GAN models are capable to reproduce both the mean and the width of the Geant4 distributions. 

\begin{figure}[ht!]  
    \centering
    \subfigure{\includegraphics[width=.45\textwidth, clip=true]{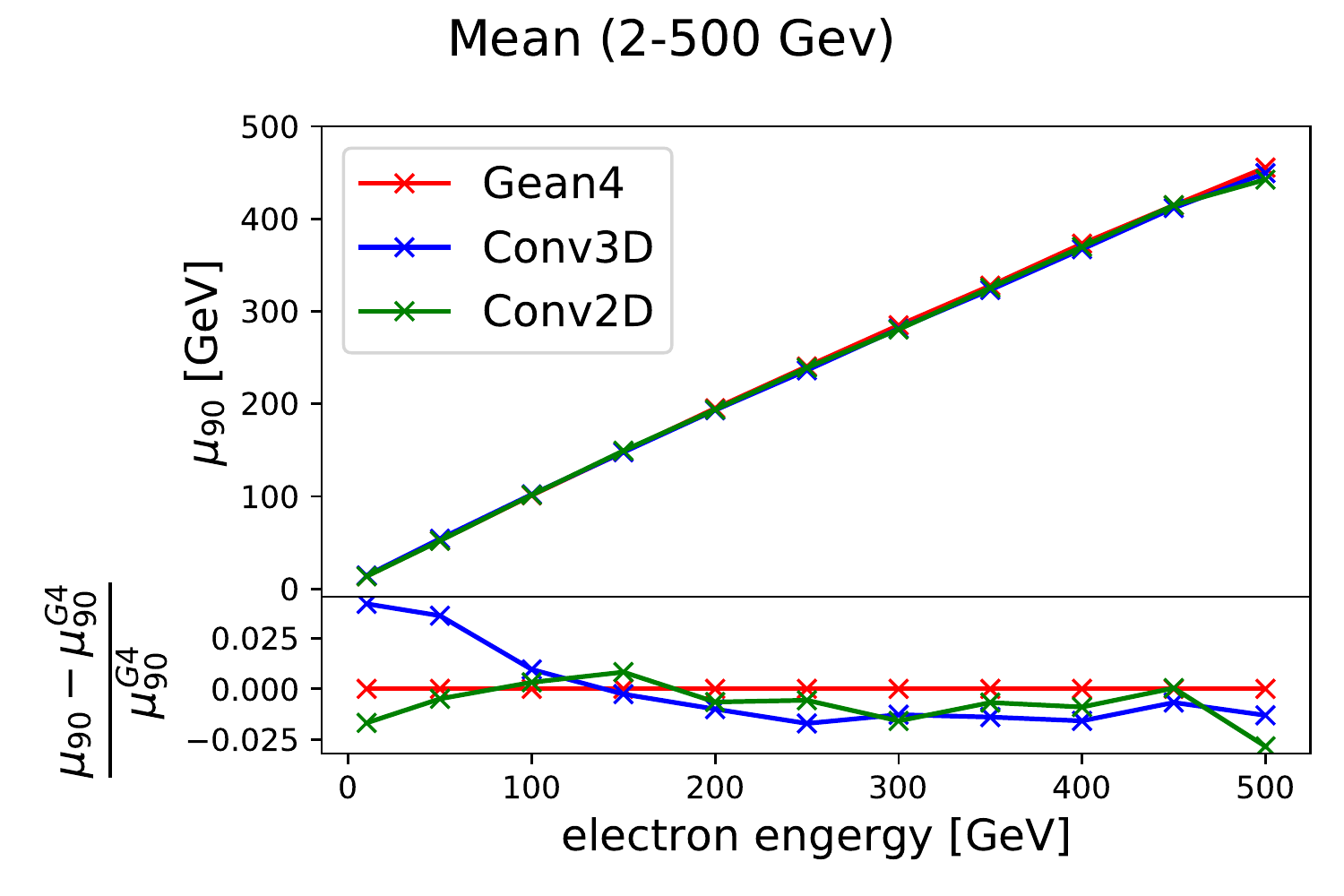}}
    \subfigure{\includegraphics[width=.45\textwidth, clip=true]{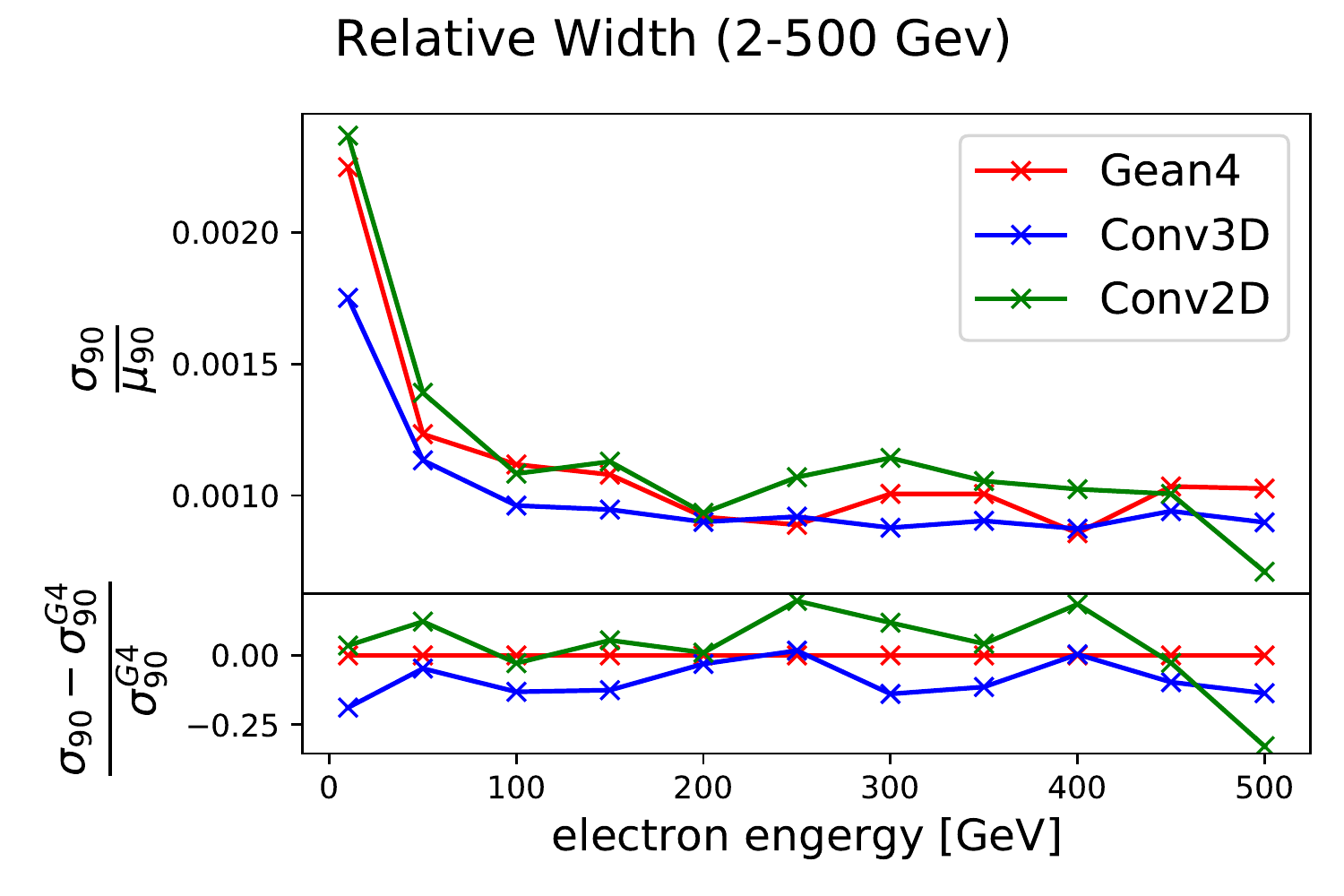}}
    \caption{Deposited energy  mean $\mu_{90}$ (left) and relative width $\sigma_{90}/\mu_{90}$ (right). The lower panels show the relative deviation to Geant4.}
    \label{fig:overall_mean}
\end{figure}

Figure \ref{fig:overall_mean} shows additional details about the deposited energy with respect to the incident electron energy. The GAN electrons are generated using the same input energy as the Geant4 events to minimize statistical effects. We calculate the root-mean-square and the standard deviation of the 90\% core of the distribution, labeled with $\mu_{90}$ and $\sigma_{90}$ respectively. Both distributions are correctly reproduced by the GANs, although a slight degradation is visible at the start and the end of the energy range. 

\section{Conclusion and Future Work}
\label{sec:conclusion}
This work compares three alternative Conv2D neural network architectures for generating 3D calorimeter shower images. We focus in the further discussion on our best Conv2D architecture (in terms of computational resources), the G1 three branches model, by comparing its performance to the Conv3D. It achieves the largest speed up ($8\,000$x) compared to Geant4 and a factor of 1.29x compared to the baseline Conv3D model. 
Additionally, our Conv2D model utilizes the GPU with a factor of 3.29x less than the Conv3D model, leaving still the possibility to achieve further acceleration by using multiple streams.
As far as detailed physics validation is concerned, the G1 Conv2D model achieves similar or better performance (in the description of the distributions tails) than the baseline Conv3D architecture developed in \cite{EnergyGAN}.

In conclusion, we exhibit how a computationally simpler model, based on two dimensional convolutions, can be used to correctly reproduce three dimensional images. The demonstrated model in this research is not a finished software product which will be employed exactly in this form. Nevertheless, a similar approach, adopted to the specific calorimeter geometry, will most presumably be used in the future HL-LHC phase to counteract the exploding demands for simulated data.
Beyond the field of High Energy Physics, this work joins the efforts ongoing in the computer vision community to design accurate models suitable for cases in which computational resources are limited.

\section*{\uppercase{Acknowledgements}}

\noindent 
This work has been sponsored by the Wolfgang Gentner Programme of the German Federal Ministry of Education and Research.

\bibliography{Chep_bib}

\end{document}